\documentclass[sn-basic,iicol]{sn-jnl}

\newcommand{\dUT}{UT1$-$UTC}

\jyear{2022}

\begin{document}

\title[Co-located GNSS and VLBI Intensive Stations: A case study of Maunakea]{The Importance of Co-located VLBI Intensive Stations and GNSS Receivers}
\subtitle{A case study of the Maunakea VLBI and GNSS stations during the 2018 Hawai`i earthquake}

\author*[1,2]{\fnm{Christopher} \sur{Dieck}}\email{christopher.dieck@navy.mil}

\author[1]{\fnm{Megan C.} \sur{Johnson}}

\author[3]{\fnm{Daniel S.} \sur{MacMillan}}

\affil[1]{\orgdiv{Celestial Reference Frame Department}, \orgname{United States Naval Observatory}, \orgaddress{\street{3450 Massachusetts Avenue NW}, \city{Washington}, \state{DC}, \postcode{20392}, \country{USA}}}

\affil[2]{\orgdiv{Department of Physics}, \orgname{The Catholic University of America}, \orgaddress{\street{620 Michigan Avenue NE}, \city{Washington}, \state{DC}, \postcode{20064}, \country{USA}}}

\affil[3]{\orgname{NVI, Inc., NASA Goddard Space Flight Center}, \orgdiv{Code 61A}, \orgaddress{\street{8800 Greenbelt Road}, \city{Greenbelt}, \state{MD}, \postcode{20771}, \country{USA}}}

\abstract{Frequent, low-latency measurements of the Earth's rotation phase, expressed as \dUT, critically support the current estimate and short-term prediction of this highly variable Earth Orientation Parameter (EOP). Very Long Baseline Interferometry (VLBI) Intensive sessions provide the required data. However, the Intensive \dUT\ measurement accuracy depends on the accuracy of numerous models, including the VLBI station position. Intensives observed with the Maunakea (Mk) and Pie Town (Pt) stations of the Very Long Baseline Array (VLBA) illustrate how a geologic event (i.e., the $M_w$ 6.9 Hawai`i Earthquake of May 4th, 2018) can cause a station displacement and an associated offset in the values of \dUT\ measured by that baseline, rendering the data from the series useless until it is corrected. Using the non-parametric Nadaraya-Watson estimator to smooth the measured \dUT\ values before and after the earthquake, we calculate the offset in the measurement to be 75.7 $\pm$ 4.6 $\mu$s. Analysis of the sensitivity of the Mk-Pt baseline's \dUT\ measurement to station position changes shows that the measured offset is consistent with the 67.2 $\pm$ 5.9 $\mu$s expected offset based on the 12.4 $\pm$ 0.6 mm total coseismic displacement of the Maunakea VLBA station determined from the displacement of the co-located global navigation satellite system (GNSS) station. GNSS station position information is known with a latency on the order of tens of hours, and thus can be used to correct the a priori position model of a co-located VLBI station such that it can continue to provide accurate measurements of the critical EOP \dUT\ as part of Intensive sessions. In the absence of a co-located GNSS receiver, the VLBI station position model would likely not be updated for several months, and a near real-time correction would not be possible. This contrast highlights the benefit of co-located GNSS and VLBI stations in support of the monitoring of \dUT\ with single baseline Intensives.}


\maketitle

\section{Introduction}
\label{intro}

Precise and current knowledge of the phase of the Earth's rotation, represented as Universal Time (UT1), is paramount to several applications, from precise pointing of astronomical telescopes to navigation with global navigation satellite systems (GNSSs) \citep{gambis_earth_2011}. However the rotation rate of the Earth is highly variable and challenging to predict. Thus, it is critical to regularly monitor UT1, expressed as \dUT, the difference between UT1 and the very regular and predictable time scale Coordinated Universal Time (UTC).

Very Long Baseline Interferometry (VLBI) is the most accurate and precise technique used to determine \dUT. Additionally, VLBI is unique in its ability to measure UT1 directly. Satellite techniques for determining Earth orientation parameters (EOPs), like GNSS, cannot differentiate between changes in the orbital parameters of the satellites and the Earth's rotation phase. To measure all five EOPs as well as contribute to the maintenance of the International Celestial Reference Frame \citep[ICRF,][]{charlot_third_2020} and the International Terrestrial Reference Frame \citep[ITRF,][]{altamimi_itrf2014_2016}, the International VLBI Service for Geodesy and Astrometry (IVS) organizes two 24-hr multi-station observing sessions each week \citep{nothnagel_international_2017}. Due to the time-consuming data transfer and the large dataset being correlated and processed, the latency of these sessions is generally 10--15 days, too long to be useful in the short-term predictions of the International Earth Rotation and Reference Frame Service (IERS) Rapid Service/Prediction Center (RS/PC). Observing sessions with latency on the order of a day or less are key to maintaining accurate and precise knowledge of the current value of \dUT\ and allowing for accurate short-term predictions \citep{luzum_improved_2010}.

VLBI sessions of short duration (1--2 hours), typically utilizing only two stations, and generally with long east-west baselines, are still sensitive to \dUT, but require sufficiently few resources that they can be observed daily and produce a \dUT\ estimate with a latency of 24 hours or less. Such sessions, called ``Intensives'', have been observed since 1984 \citep{robertson_daily_1985}, and several Intensive series are observed at the present time \citep[i.e.,][]{finkelstein_eop_2011, nothnagel_international_2017}.

With the geometric limitations of Intensives and the relatively low number of observations in them, only a few parameters can be estimated. To account for the physical realities of the observing system that cannot be estimated, analysts rely on accurate models to provide the a priori information needed to make an accurate measurement of \dUT. \citet{nothnagel_impact_2008}, \citet{malkin_impact_2011, malkin_impact_2013}, \citet{nilsson_improving_2017}, \citet{landskron_improving_2019}, and forthcoming work by \citet{kern_importance_2022} discuss how errors in the a priori models (e.g., nutation, polar motion, tropospheric gradients, ocean tide loading, etc.) propagate into errors in the estimated value of \dUT.

Beyond those small errors, earthquakes and other geologic activity can cause large departures from the typical station position motion in the a priori models. The position motion models for VLBI stations are updated no more frequently than every few months, and require that several 24-hr sessions that include the affected station have been observed subsequent to the geologic event. Thus, in the absence of means to update the station position in near-real-time, a station affected in such a way cannot be used to provide the desired accurate, low-latency measurement of \dUT. \citet{macmillan_effects_2012} demonstrated that a VLBI station position model can be modified by incorporating station position measurements from a co-located GNSS receiver. Applied with sufficiently low latency, these corrections can enable a station to continue to participate in Intensive sessions and provide accurate measurements of \dUT.

In this work, we investigate the impact of the Hawai`i earthquake of May 4, 2018 (moment magnitude ($M_w$) 6.9) on geodetic stations close to the earthquake's epicenter. Roughly 80 km from the epicenter, near the summit of Maunakea, sit the MKEA GNSS station and the MK-VLBA (Mk) VLBI station of the Very Long Baseline Array (VLBA), less than 90 m apart. The MK-VLBA station participated in a daily Intensive series with the Pie Town VLBA Station (PIETOWN; Pt) from 2011--2021 \citep{geiger_intensifying_2019}. The occurrence of an earthquake in relatively close proximity to both a GNSS station and a VLBI station that is regularly involved in an Intensive series provides an opportune case study for exploring the connection between a coseismic displacement of the GNSS station and any concurrent discontinuity in the value of \dUT\ measured by the VLBI Intensive series. A critical part of that connection is the sensitivity of the VLBI Intensive \dUT\ measurement to changes in the positions of the participating stations. That is explored here by simulating small errors in the a priori VLBI station positions and calculating the resulting offsets in the measurements of \dUT.

The details of the Mk-Pt data series are given in Section \ref{Observations} along with the description of their co-located GNSS receivers. The IVS Intensive series observed between the 20m radio telescopes of the K\=oke`e Park Geophysical Observatory (KOKEE; Kk) and the Wettzell Geodetic Observatory (WETTZELL; Wz) is included as a control series in this study and is also described in Section \ref{Observations}. Section \ref{Analysis} contains the description of the analysis techniques, particularly the data smoothing that enables the estimation of discontinuities and the sensitivity analysis, as well as the results of the analysis. These results are discussed in Section \ref{Discussion}. Section \ref{Conclusion} concludes the work.

\section{Observations}
\label{Observations}

\subsection{K\=oke`e--Wettzell VLBI Intensives}
\label{Obs:KkWz}

\begin{figure}[ht]
  \centering
  \includegraphics[width=76mm]{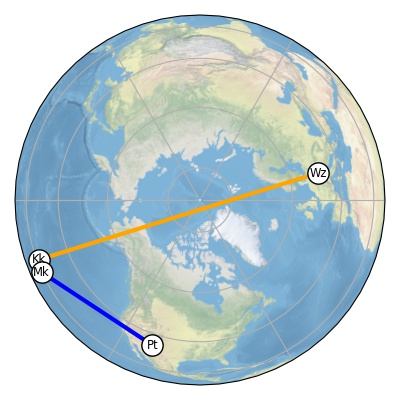}
\caption{Map showing the locations of the VLBI stations used in the IVS Kk-Wz and VLBA Mk-Pt Intensives.}
\label{fig:baselinemap}
\end{figure}

The VLBI stations KOKEE on the island of Kaua`i in Hawaii, USA ($\sim$500 km to the northwest of the MK-VLBA station on the island of Hawai`i) and WETTZELL in Germany together have participated in Intensives under the auspices of the IVS for over 20 years. These sessions are designated as the ``INT1''s, and are typically observed Monday through Friday at 1830 UTC. The Kk station is in the northern tropics (22.126$^{\circ}$ N) and the Wz station is in the mid-latitudes of the northern hemisphere (49.145$^{\circ}$ N). They have a longitude difference of 172.5$^{\circ}$ and a baseline length (the straight-line distance between the two stations) of 10357 km. Figure \ref{fig:baselinemap} shows the locations of the two stations. Simulations performed by \citet{schartner_optimal_2021} and \citet{kern_importance_2022} indicate that the length and orientation of this baseline is close to the optimal baseline for Intensives that could be achieved from the Earth's surface.

\begin{table}[ht]
\begin{center}
\begin{minipage}{76mm}
\caption{Properties of the Kk-Wz series and its observational setup.}
\label{tab:KkWz}
\begin{tabular}{@{}l|l@{}}
\toprule
Property & Value \\
\midrule
\# of Channels     & 16 \\
Channel Bandwidth  & 4 MHz \\
Sampling Rate      & 1 bit/sample \\
Total Data Rate    & 128 Mbps \\
\midrule
S-Band Frequencies & 2212.99, 2222.99, 2257.99, \\
\ \ (lower edge)   & 2297.99, 2317.99, 2322.99 MHz \\
X-Band Frequencies & 8178.99, 8182.99, 8222.99, 8422.99, \\
\ \ (lower edge)   & 8562.99, 8682.99, 8782.99, 8842.99, \\
                   & 8858.99, 8862.99 MHz \\
\midrule
Telescope Diameters & 20 m \\
Baseline Length     & 10357 km \\
\botrule
\end{tabular}
\end{minipage}
\end{center}
\end{table}

The Kk-Wz IVS Intensive sessions observe sources in the S-band (2.2-2.4 GHz) and X-band (8.1-8.9 GHz) in right circular polarization. To enable the calibration of the ionospheric delay for each observation, the radio signal is simultaneously passed into both receivers. The data is channelized into 6 channels in the S-band and 10 in the X-band, for a total of 16 channels each with a 4 MHz bandwidth (increased to 8 MHz bandwidth per channel on April 1, 2022). The data are then sampled at the Nyquist rate and digitized at 1 bit/sample for a total data rate of 128 Mbps (256 Mbps after channel bandwidth increase). These recording details, the frequencies of the channels, and other details of the sessions are listed in Table \ref{tab:KkWz}.

Like most VLBI Intensive sessions, the IVS INT1s are 1-hour in duration. The session schedules are created using the program \texttt{sked} \footnote{\url{https://ivscc.gsfc.nasa.gov/IVS_AC/sked_cat/SkedManual_v2018October12.pdf}} using a mode which calculates the duration of each scan based on the expected flux density of the source and a user supplied signal to noise ratio (SNR) target. These schedules can include up to $\sim$20 unique observations of radio quasars called ``scans''.

\subsection{Maunakea---Pie Town VLBI Intensives}
\label{Obs:MkPt}
In late 2011, the United States Naval Observatory (USNO) began observing Intensive sessions on the Very Long Baseline Array (VLBA). Though the naive optimal choice for an Intensive baseline on the VLBA would be the longest baseline, which is the 8611 km baseline between the Maunakea and St. Croix (SC-VLBA; Sc) stations, simulations by \citet{kern_simulation_2021} identify the baseline between the Maunakea and Hancock (HN-VLBA; Hn) stations as the baseline with the lowest expected formal errors. However, when the series was initiated, practicalities led to the consideration of other stations of the array for use in the Intensive series. One of the challenges with Intensives is estimating the water content of the atmosphere, and very moist sites are likely to have elevated measurement uncertainties or systematic errors. The Maunakea station is at an elevation of 3763 m, thus above much of the atmospheric moisture, and Pie Town (PIETOWN; Pt) is in a semiarid climate at an elevation of 2365 m above sea level. Both stations also had access to fiber optic networks to enable the rapid e-transfer of data, which other stations of the VLBA did not have when the series commenced. The atmospheric water content of the St. Croix site, along with the desire to transfer the observation data over the internet rather than by parcel service, led to the selection of the Mk and Pt stations for use in the USNO VLBA Intensive series. The geographic locations of these stations are shown in Figure \ref{fig:baselinemap}.

\begin{table}[ht]
\begin{center}
\begin{minipage}{76mm}
\caption{Properties of the Mk-Pt series and its observational setup. Note that there are two S-Band frequency setups. The original setup was replaced by the newer setup on August 1, 2020.}
\label{tab:MkPt}
\begin{tabular}{@{}l|l@{}}
\toprule
Property & Value \\
\midrule
\# of Channels     & 16 \\
Channel Bandwidth  & 32 MHz \\
Sampling Rate      & 2 bits/sample \\
Total Data Rate    & 2048 Mbps \\
\midrule
S-Band Frequencies & 2052, 2084, 2116, \\
\ \ (lower edge), 1st setup & 2212, 2244, 2276 MHz \\ 
S-Band Frequencies & 2188, 2220, 2252, \\
\ \ (lower edge), 2nd setup & 2284, 2348, 2380 MHz \\ 
X-Band Frequencies & 8428, 8460, 8492, 8556, \\
\ \ (lower edge)   & 8620, 8684, 8748, 8812, \\
                   & 8844, 8876 MHz \\
\midrule
Telescope Diameters & 25 m \\
Baseline Length     & 4796 km \\
\botrule
\end{tabular}
\end{minipage}
\end{center}
\end{table}

Just like the IVS Kk-Wz Intensives, the Mk-Pt sessions utilized right circularly polarized simultaneous S-band and X-band observations made possible by a dichroic screen, which allows for calibration of the differential ionospheric delay for each observation. Also similarly, six channels were recorded in the S-band and ten channels in the X-band. One major difference between the IVS and VLBA stations is the backend recording hardware. The VLBA Intensives were observed with the polyphase filter bank (PFB) personality of the ROACH (Reconfigurable Open Architecture Computing Hardware) Digital Backend (RDBE). This personality provides 16 channels, each with 32 MHz bandwidth, sampled at the Nyquist rate, and digitized at 2 bits per sample for a total data rate of 2048 Mbps, 16 times more than the Kk-Wz series. The specific channel frequencies are described in Table \ref{tab:MkPt} along with other details of the sessions.

During the initial commissioning period of the VLBA Mk-Pt Intensives, the sessions were 40 minutes in duration and provided numbers of scans in the low 20s. The desired signal-to-noise ratio (SNR) could be achieved for most sources with a short scan duration of 16 seconds because of the higher data rate and higher sensitivity of the larger 25m antennas of the VLBA, compared to the Kk-Wz Intensives. These shorter scan times allowed the VLBA Intensives to observe at least as many scans as the IVS INT1s in less time. On February 15, 2013, the session duration was extended to 45 minutes, enabling the recording of closer to 30 scans per session.

There were several additional changes to the Mk-Pt sessions over the years. The duration of sessions was doubled to 90 minutes on February 24, 2017. On August 1, 2020, the duration changed again to 60 minutes, consistent with the 1-hour duration of the Kk-Wz Intensives, and the S-band frequencies were modified at that time as well to avoid persistent and worsening radio frequency interference. Beginning on December 15, 2020, the sessions were scheduled using the \texttt{VieSched++} \citep{schartner_viesched_2019} software instead of the \texttt{SCHED} \footnote{\url{http://www.aoc.nrao.edu/software/sched/index.html}} program of the National Radio Astronomy Observatory (NRAO). This change was made so that the scan length could be calculated based on the target SNR and the expected flux density of the source, rather than the fixed scan duration that \texttt{SCHED} required. After many years of observations, analysis of the Mk-Pt Intensive series showed that it suffered from unacceptably high contributions from uncorrected systematic errors (discussed in Section \ref{Disc:MkPtCharacteristics}), and the series was discontinued on April 29, 2021.

\subsection{Maunakea and Pie Town GNSS data}
\label{Obs:GNSS}
The International GNSS Service (IGS) collects data from over 500 GNSS stations \citep{johnston_international_2017}. The IGS stations are permanently mounted GNSS antennas and receivers that operate continuously and utilize one or more satellite system(s) to determine their position with a precision of $\sim$3 mm in the horizontal and $\sim$6 mm in the vertical \citep{johnston_international_2017}. One of these stations is co-located with each of the MK-VLBA and PIETOWN VLBA stations, and are identified as MKEA and PIE1, respectively. The GNSS and VLBI stations are in close proximity to each other (within 90 m) and are effectively in the same place, geologically. Therefore, tectonic plate motion and site movement can be assumed to be the same for both stations at each site. Additionally, the impact of sources of systematic effects on the measurements of the position of each station, such as solid earth tides and the troposphere, will be almost identical at each site. Under these assumptions, the measured position evolution of one station at a site determined through one technique can be transferred to the other station at that site. In this instance, it particularly allows us to determine if and which of the MK-VLBA and PIETOWN stations moved, and by how much, potentially causing the \dUT\ discontinuity.

\begin{table}[ht]
\begin{center}
\begin{minipage}{76mm}
\caption{Properties of the MKEA GNSS station.}
\label{tab:MKEA}
\begin{tabular}{@{}l|l@{}}
\toprule
Property & Value \\
\midrule
Site code           & MKEA00USA \\
Distance to MK-VLBA &  87.772 m \\
\midrule
Antenna Type & Javad RingAnt-DM \\
Radome       & SCIS \\
Receiver Type   & JAVAD TRE\_3GTH DELTA \\
\botrule
\end{tabular}
\end{minipage}
\end{center}
\end{table}

\begin{table}[ht]
\begin{center}
\begin{minipage}{76mm}
\caption{Properties of the PIE1 GNSS station.}
\label{tab:PIE1}
\begin{tabular}{@{}l|l@{}}
\toprule
Property & Value \\
\midrule
Site code           & PIE100USA \\
Distance to PIETOWN &  61.795 m \\
\midrule
Antenna Type & Ashtech 701945 Rev. E \\
             & Choke-ring \\
Radome       & None \\
Receiver Type   & JAVAD TRE\_3GTH DELTA \\
\botrule
\end{tabular}
\end{minipage}
\end{center}
\end{table}

The receivers for the GNSS stations are changed or altered periodically, which can cause discontinuities in their position histories. In this analysis, for each station, we opt to utilize only data from the same equipment setup, choosing the time period that spanned the moment of the \dUT\ discontinuity. According to the station log for MKEA \footnote{\url{https://files.igs.org/pub/station/log_9char/mkea00usa_20220428.log}}, this time frame is from February 23, 2016 through September 23, 2018. Details of the MKEA station are given in Table \ref{tab:MKEA}. From the PIE1 log \footnote{\url{https://files.igs.org/pub/station/log_9char/pie100usa_20210802.log}}, this time frame is from June 30, 2017 through October 1, 2018, and properties of this station are described in Table \ref{tab:PIE1}. The time period where both stations have unchanged setups is thus June 30, 2017 through September 23, 2018.

The IGS produces a daily product that contains the position of each of the stations in its network. For the ranges of days described, we retrieve the daily combined terrestrial reference frame solutions from the 3rd IGS reprocessing campaign (``Repro3'') \footnote{\url{https://cddis.nasa.gov/archive/gnss/products}}, and extract the positions for the MKEA and PIE1 stations. The processing method for the Repro3 data is described on the IGS website\footnote{\url{https://igs.org/acc/reprocessing/\#repro3-conventions-modelling}}.

\section{Analysis and Results}
\label{Analysis}

\subsection{VLBI}
\label{Anl:VLBI}

\subsubsection{VLBI Session Editing and \dUT\ Calculation}
\label{Anl:VLBI:Editing}
The purpose of VLBI Intensives is to measure \dUT. To extract that value, the data of each session are first correlated and processed by the correlator assigned to the session. For the Mk-Pt VLBA Intensives and the Kk-Wz IVS Intensives, this is the Washington Correlator based at the USNO. The USNO VLBI Analysis Center then updates the delay model and associates meteorological data with the session to determine the observed group delays. Then, through interactive session editing with the \texttt{nuSolve} software package \citep{bolotin_first_2012}, any delay ambiguities are resolved, and a group delay solution is fit at both the X- and S-bands. Corrections to the group delay due to the ionosphere are then calculated and applied, followed by iterative inflation of the observation uncertainty (so that the reduced $\chi^2$ of the delay residuals with respect to the delay model is unity) and outlier elimination.

To calculate the value of \dUT\ at each session, a parameterized model is fit to the data using a linear least squares minimization fit. The model contains a single \dUT\ parameter, two or three clock polynomial parameters for the non-reference station(s), and a zenith wet delay for each station. For a two-station Intensive, this means that five or six parameters (depending on the number of clock parameters) are estimated for each session. This is done with the \texttt{Calc/Solve} package \citep{ryan_mark_1980, ma_radio-source_1986, caprette_crustal_1990, ryan_nasa_1993}. Station positions and \dUT\ cannot be estimated simultaneously from observations from a single baseline. To accurately estimate \dUT, the station positions and velocities must be fixed to correct values. There are numerous other values that are fixed in the model. The extragalactic radio source positions are assumed to be correct and the X and Y polar motion EOPs are interpolated from the IERS RS/PC's products. Models describing the celestial pole motion and alterations to the station positions due to geophysical phenomena are assumed following the recommendations of the IERS Conventions \citep{petit_iers_2010}. From the least squares fit a measurement of \dUT\ and a formal error are produced for each viable session. Periodically, the values are re-calculated as part of a self-consistent global solution of 24-hour VLBI sessions that simultaneously solves for source positions, station positions, station velocities, and EOPs. The \dUT\ values used in this analysis are from the \texttt{usn2019c} Intensive solution \footnote{\url{https://cddis.nasa.gov/archive/vlbi/ivsproducts/eopi/usn2019c.eopi.gz}}. Included in that data set are 2238 \dUT\ measurements in the Mk-Pt series and 1629 measurements in the Kk-Wz series over the same time period spanned by the Mk-Pt sessions (November 4, 2011--January 1, 2020).

\subsubsection{\dUT\ Residuals and the Nadaraya-Watson Estimator}
\label{Anl:VLBI:NWE}
Before a time series of \dUT\ measurements can be used in a combination with other measurements of \dUT, it must be evaluated for its accuracy and precision by comparing the series to a reference series. A reference series takes multiple inputs from multiple measurement techniques and combines them to create a time series that ostensibly more accurately reflects the true EOPs than any one contributing series. Three institutions produce regularly updated reference series: the Observatoire de Paris (OPAR) \footnote{\url{https://hpiers.obspm.fr/eoppc/eop/eopc04/eopc04_IAU2000.62-now}}, the US Naval Observatory (USNO) \footnote{\url{https://maia.usno.navy.mil/ser7/finals2000A.all}}, and NASA's Jet Propulsion Laboratory (JPL) \footnote{\url{https://keof.jpl.nasa.gov/predictions/latest_midnight.eop}}. In all three of the reference series, the Mk-Pt series is not used as an input series while the Kk-Wz series is. The analysis presented here uses the \texttt{finals2000A.all} file produced by the USNO as the reference series.

Comparison of an observation series to the reference series consists of calculating the residuals between the two series. Though the reference series is reported at a consistent time of day, midnight UTC, the entries in the observation series are not at the same time as the reference series, nor always at a consistent time of day. So, the reference series is interpolated to the epoch of each VLBI \dUT\ measurement using a Lagrange interpolation of order two before subtracting the reference series from the observation series. The formal error of each resulting residual is the quadrature sum of the UT1$-$UTC formal error and the formal error of the interpolated reference series value. Any systematic offset or drift in the residual series is removed by subtracting a line fitted to the data using a weighted linear least squares fit, where the weights are the inverse square of the formal errors of the UT1$-$UTC residuals. If the corrected residuals are relatively flat and free from anomalies, the series can be considered for inclusion in the creation of a reference series.

\begin{figure*}[ht]
\centering
  \includegraphics[width=\textwidth]{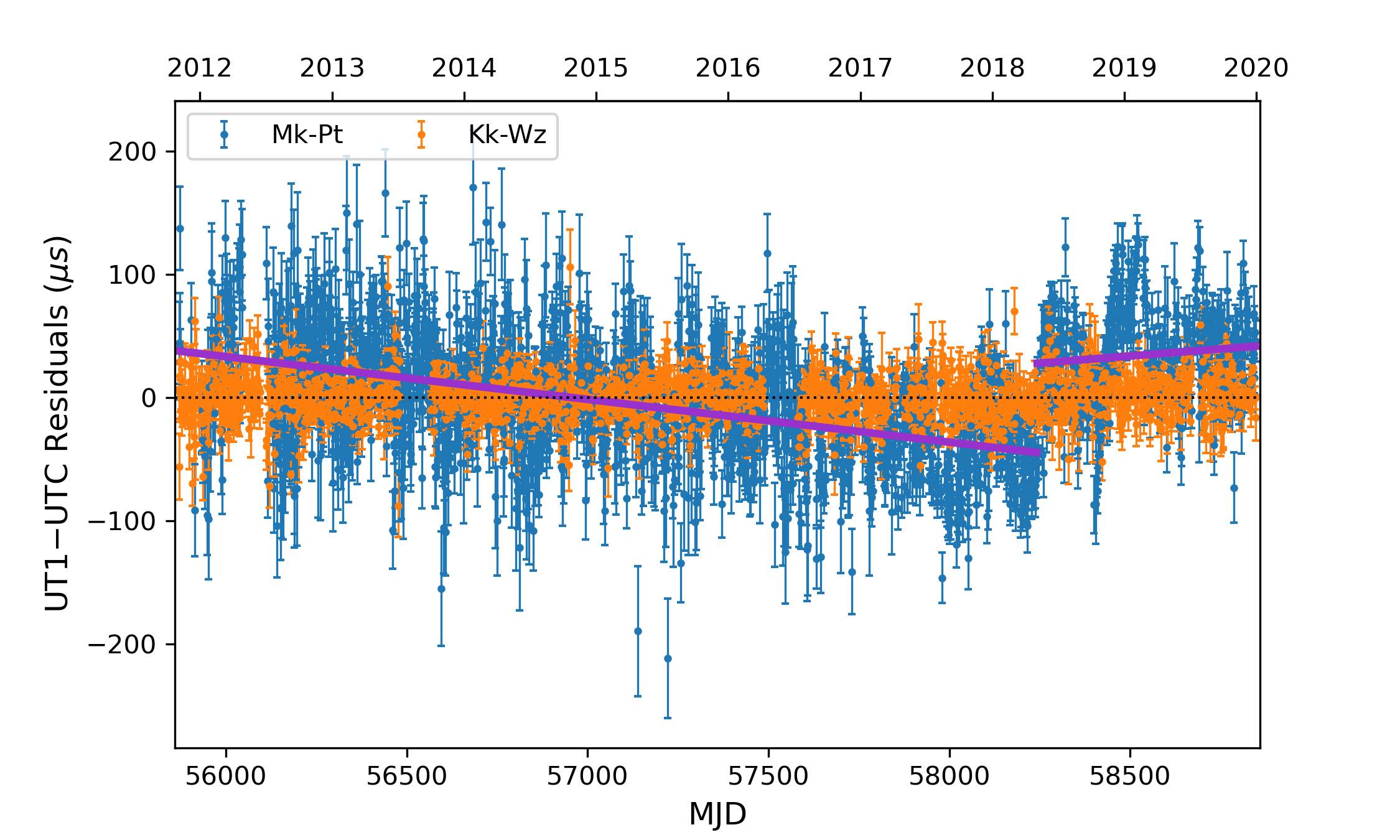}
\caption{The residuals of \dUT\ measurements as compared to the USNO reference series for the Mk-Pt VLBA Intensives, in blue, and the Kk-Wz IVS Intensives, in orange, as a function of time, denoted in modified Julian date (MJD) and Gregorian year. Residuals with large formal errors have been removed from both series ($> 54.1\ \mu$s for Mk-Pt and $> 34.4\ \mu$s for Kk-Wz) as have large outliers. The systematic offset and drift calculated from the entirety of each series has also been subtracted. Even after these treatments, the Mk-Pt series exhibits a negative slope from 2012 through mid-2018, while from mid-2018 to 2020 the slope is relatively flat and the residuals have a positive offset, as indicated by the purple lines.}
\label{fig:resids}
\end{figure*}

Figure \ref{fig:resids} shows the modified residuals for both the Kk-Wz and Mk-Pt observation series, and a sharp discontinuity in the Mk-Pt residuals is present in mid-2018 while there is no such discontinuity in the Kk-Wz series. The \dUT\ residuals must be continuous with time, so a series containing a discontinuity does not make for a suitable observation series. Before the Mk-Pt series can be used in a combined series, the reason for the discontinuity in the residuals must be determined and rectified. The first step in doing that is to determine the magnitude of the offset. However, doing so is not as simple as calculating the difference in the values before and after the jump. It is clear that there is scatter around the mean residual, so first we need to establish that mean value. However, there is structure in the residuals with time, and a mean over all time would not appropriately capture that variation. A non-parametric regression providing an estimate of the local mean about a point, such as the Nadaraya-Watson estimator \citep[NWE,][]{nadaraya_estimating_1964, watson_smooth_1964}, is suitable for this circumstance.

The NWE is a function that returns an estimate of the value of a regression function $m$ at a given point, $x$, based on an average of a sample $\{(X_i, Y_i)\}^N_{i=1}$, weighted by a kernel, $K$. The original NWE does not take into account the uncertainty in the sample values. The sample of interest here does have formal errors on the \dUT\ values, and that additional information should not be ignored. Thus, in this work, the kernel used with the basic NWE is extended so that the weights are determined by the product of a standard Gaussian kernel and the inverse square of the formal error, $\sigma_{Y_i}$. Specifically, for each point $x$ for which an estimate of the local mean is desired,
\begin{align}
    \widehat{m_x} &= \sum_{i=1}^{N} w_i Y_i \nonumber \\
        &= \sum_{i=1}^{N}\frac{K(x - X_i)}{\sum_{j=1}^{N} K(x - X_j)} Y_i
\end{align}
where $N$ is the number of samples in the series and $K$ is the weighted Gaussian kernel, and given as
\begin{equation}
    K(x - X_i) = \frac{1}{\sigma_{Y_i}^2} \exp\left[-\left(\frac{x - X_i}{\sqrt{2}h}\right)^2\right]
\end{equation}
with $h$ as the kernel bandwidth.

In this work, the uncertainties of the estimates are calculated via the propagation of uncertainty as
\begin{equation}
\label{eqn:nwe_uncertainty}
    \sigma_{\widehat{m_x}} = \sqrt{\sum_{i=1}^{N}\frac{[\sigma_{Y_i} K(x - X_i)]^2}{[\sum_{j=1}^{N} K(x - X_j)]^2}}.
\end{equation}
A bootstrap resampling approach to estimating the confidence interval of the regression function could also be considered.

The kernel bandwidth is determined through leave-one-out cross-validation. In this process the NWE is applied to find the estimated value at each $X_i$ of the $N$ samples while excluding that sample from the data used to calculate the estimate. The statistic
\begin{equation}
    S = \frac{1}{N} \sum_{i=1}^{N} \lvert Y_i - \widehat{m_{X_i}} \rvert
\end{equation}
is used to evaluate how well the estimated values fit the sample data. This statistic is calculated for a range of potential values of the kernel bandwidth, $h$, and a parabola is fit to the resulting values of the statistic in the vicinity of the minimum statistic. The bandwidth at the minimum of the parabola is the one that best fits the sample data. This critical bandwidth is subsequently used to calculate the final estimates of the regression function.

\begin{figure*}[ht]
\centering
  \includegraphics[width=\textwidth]{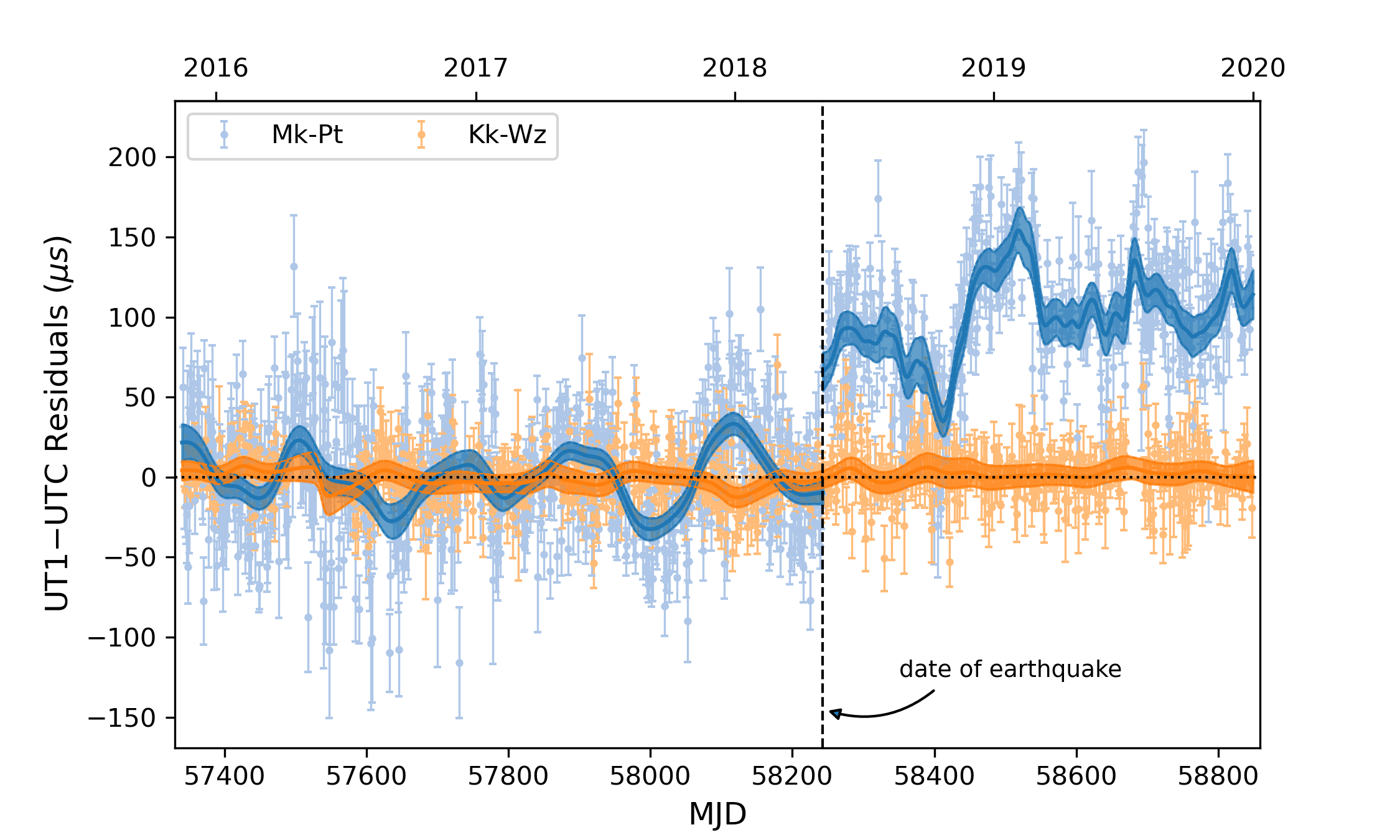}
\caption{The residuals of \dUT\ measurements as compared to the USNO reference series for the Mk-Pt VLBA Intensives, in light blue, and the Kk-Wz IVS Intensives, in light orange, as a function of time, denoted in modified Julian date (MJD) and Gregorian year. The Mk-Pt series is split into two sections, one on either side of May 4, 2018 (the date of the seismic event experienced at the MK-VLBA station). Both series have had large outliers and any systematic offset and drift removed, though for the Mk-Pt series the systematic corrections were calculated for the first section only and applied to both. In corresponding darker colors, the residuals are overlaid with the regression functions calculated by the Nadaraya-Watson estimator with 3-sigma confidence intervals shown.}
\label{fig:doublefit}
\end{figure*}

Before the NWE is applied to each series, individual UT1$-$UTC residuals with large formal errors are removed. The applied threshold is calculated for each series as the mean formal error of the residuals plus three times the standard deviation of those formal errors (54.1 $\mu$s for Mk-Pt and 34.4 $\mu$s for Kk-Wz). Figure \ref{fig:doublefit} shows the NWE fit to both data series in the timeframe from after the major maintenance at the MK-VLBA station of November 14, 2015 to January 1, 2020. In this range, there are 1142 Mk-Pt sessions and 764 Kk-Wz sessions. The regression function is estimated at the times of the Intensive measurements and applied to the Mk-Pt series in two sections, one on either side of the discontinuity, while the Kk-Wz series is fit in a single section. After the NWE is applied, outliers are also removed. An outlier is defined as a data point that has a absolute normalized residual greater than a specified value. Here the residual is the difference between the UT1$-$UTC residual and the estimated value of the UT1$-$UTC residual calculated by the NWE. That post-estimate residual of the UT1$-$UTC residual is normalized by the uncertainty in the post-estimate residual, which is calculated as the quadrature sum of the uncertainty of the UT1$-$UTC residual and the uncertainty in the Nadaraya-Watson estimate as calculated by Equation \ref{eqn:nwe_uncertainty}. These outliers are removed iteratively, with the NWE re-calculated after every removal of outliers until no more are removed.

\subsubsection{Mk-Pt \dUT\ Discontinuity Estimate}
\label{Anl:VLBI:MkPt_disp}
Now equipped with a method of estimating the local mean value of the residuals, we can calculate the difference of the \dUT\ values from before and after the observed discontinuity in the Mk-Pt series. When preparing this calculation, it became clear that the value of the estimated jump is fairly sensitive to the kernel bandwidth used in the estimation. In turn, the kernel bandwidth, determined through leave-one-out cross-validation, depends on how many and which points are present in the time series. A stringent outlier threshold of three times the normalized residual reduces the number of points used in the estimator, which results in a small bandwidth for the second section of the Mk-Pt data series. This makes the regression function very sensitive to high frequency variations in the UT1$-$UTC residuals, decreasing the benefit of the fit for the second section. On the other hand, a generous outlier threshold of five times the normalized residual removes very few points, and the calculated bandwidth for the second part is more similar to the bandwidth calculated for the first section.

An outlier threshold of four times the normalized residual of a data point provides a compromise. It is small enough that the visually obvious outliers are eliminated, yet doesn't make it so that the kernel bandwidth is so small as to render the fit unhelpful. Thus, this is the value used in preparing the data series for the final analysis shown in Figure \ref{fig:doublefit}. From those non-parametric regression functions, where the critical kernel bandwidths for the Mk-Pt series are 17.240 and 6.629 for before and after the earthquake, respectively, and 18.281 for the Kk-Wz series, the shift in the estimate of \dUT\ is
\begin{equation}
    \label{eqn:MkPt_dUT1_shift}
    \Delta(\text{\dUT})_{MK-VLBA} = 75.7 \pm 4.6\ \mu s.
\end{equation}
This value is in the middle of the range of values calculated with different outlier thresholds, so is deemed a reasonable value with which to move forward, understanding that the calculated uncertainty likely under-represents the true uncertainty in the calculated difference.

\subsection{GNSS}
\label{Anl:GNSS}

\begin{figure*}[ht]
\centering
  \includegraphics[width=\textwidth]{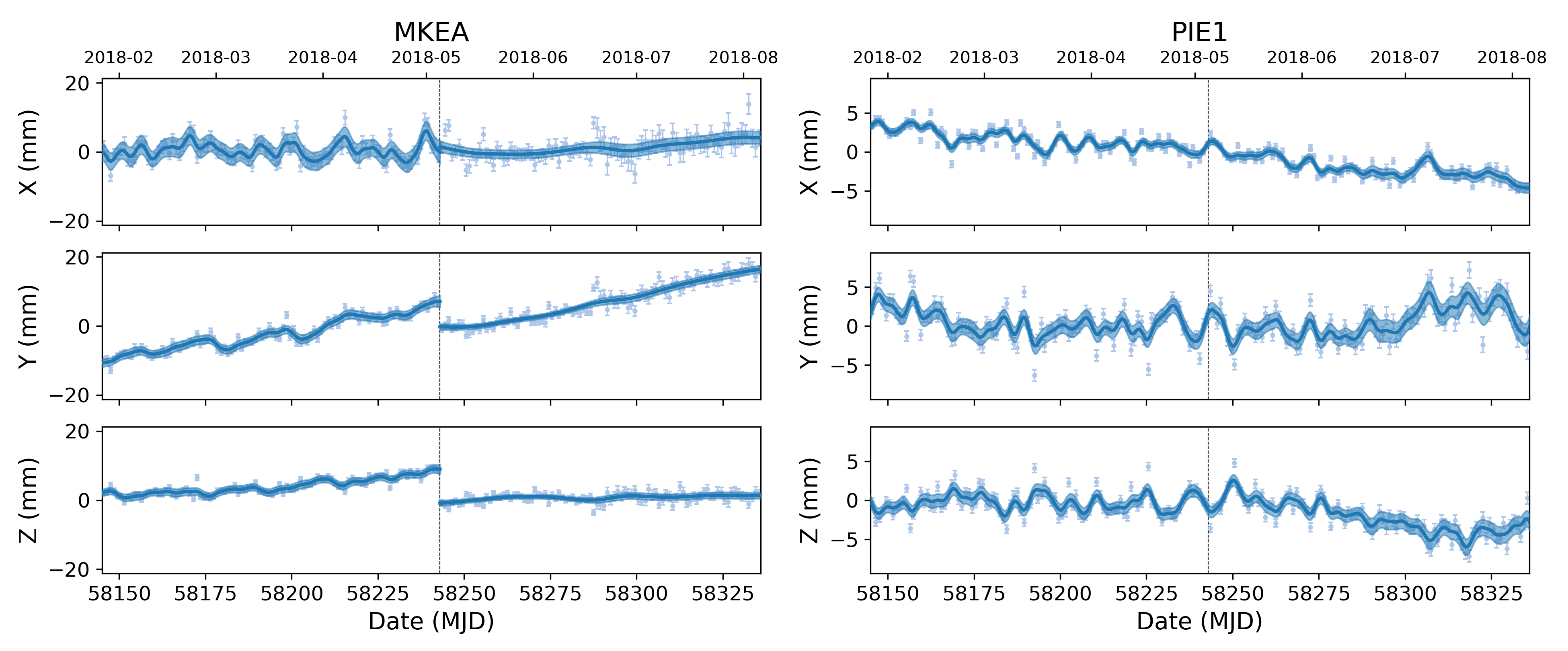}
\caption{The IGS Repro3 time series of the X, Y, and Z positions of the MKEA and PIE1 GNSS stations co-located with the MK-VLBA and PIETOWN VLBA stations, respectively. For each coordinate series the mean position as determined from the measurements from May 5th through May 25th, 2018 inclusive is subtracted to more easily see the relative changes. The date of the earthquake is marked by the vertical dotted line in both stations. The solid blue line is the regression function calculated with the Nadaraya-Watson estimator, with 3$\sigma$ confidence bands. Though the regression function was calculated over the time span identified in Section \ref{Obs:GNSS}, June 30, 2017--September 23, 2018, the plot shows only the three months on either side of the earthquake so that the discontinuity is more visible. A position shift is evident in the MKEA data at the time of the earthquake, so the smoothing is applied to the MKEA data in two parts. For the PIE1 data, which does not show a position discontinuity, the smoothing is done in one part. Note the difference in the scale of the vertical axis between the two stations, which enhances the variability of the PIE1 series relative to the MKEA series.}
\label{fig:IGS}
\end{figure*}

The purpose of investigating the position history of the GNSS stations co-located with the VLBA stations is to see if a discontinuity in that time series also occurred in the vicinity of the seismic event that is suspected of causing the discontinuity in \dUT. Visually inspecting IGS Repro3 station position data from approximately three months on either side of the seismic event, as shown in Figure \ref{fig:IGS}, a clear break is present in the MKEA position time series while no break is present at PIE1.

As with the VLBI measurements of \dUT, determining a position shift is not as simple as taking the difference in the positions from before and after the seismic event. So, similar to the analysis of the \dUT\ data from the Mk-Pt baseline, we apply the NWE to the MKEA position history to identify estimated positions of the GNSS station from immediately prior to and immediately after the seismic event \footnote{The station log for MKEA reports that 3 cm of vertical station displacement was observed when a cable was replaced in December 2019. This is attributed to a slow signal degradation from the time that the cable was installed in 2015. Over a 4-5 year span, the day-to-day change in the station position values due to this effect is well below the measurement uncertainties. Because this effect wouldn't impact the coseismic displacement estimates, it is ignored in this analysis.}. The NWE is also applied to the PIE1 data, but in one continuous segment because no discontinuity is apparent. The time span of data over which the regression function is calculated is the span determined in Section \ref{Obs:GNSS}, June 30, 2017--September 23, 2018. For the GNSS kernel regression, the critical bandwidth, shown in Table \ref{tab:GNSS_kb}, determined through leave-one-out cross-validation, as was done in Section \ref{Anl:VLBI:NWE}, differed between the different coordinate axes and between the two segments of the same coordinates of MKEA.

\begin{table}[h]
\begin{center}
\begin{minipage}{76mm}
\caption{The critical kernel bandwidths, $h$, determined through leave-one-out cross-validation, of the Nadaraya-Watson Estimator applied to each of the three spatial axes for the MKEA station before and after the earthquake and the PIE1 station.}
\label{tab:GNSS_kb}
\begin{tabular}{@{}l|c|c|c@{}}
\toprule
Axis &  MKEA  pre-EQ &  MKEA post-EQ & PIE1 \\
\midrule
X & 1.134 & 5.855 & 1.311 \\
Y & 1.663 & 5.771 & 1.300 \\
Z & 1.488 & 4.870 & 1.270\\
\botrule
\end{tabular}
\end{minipage}
\end{center}
\end{table}

The resulting estimated jumps in the MKEA station position for each coordinate are listed in Table \ref{tab:MKEA_disp}. Converting the Earth Centered Earth Fixed (ECEF) coordinate system displacements into the local tangential coordinate system, the total displacement is 12.4 $\pm$ 0.6 mm down and to the south east. 

\begin{table}[h]
\begin{center}
\begin{minipage}{76mm}
\caption{MKEA station displacements, in millimeters (mm), for each orthogonal axis and the corresponding contribution to the total expected shift in \dUT\ as calculated in Section \ref{Anl:EstimatedShift}. Also tabulated is the total magnitude of the station shift and corresponding impact on the VLBI measurement of \dUT.}
\label{tab:MKEA_disp}
\begin{tabular}{@{}l|c|c@{}}
\toprule
Axis & Displacement & $\Delta$(\dUT)\ Contribution\\
\midrule
X & \hphantom{-}1.6 $\pm$ 1.2 mm & \hphantom{2}-0.8 $\pm$ 1.0 $\mu$s\\
Y &            -7.4 $\pm$ 0.6 mm & \hphantom{-}24.4 $\pm$ 3.0 $\mu$s\\
Z &            -9.9 $\pm$ 0.5 mm & \hphantom{-}43.6 $\pm$ 5.0 $\mu$s\\
\midrule
Total &        12.4 $\pm$ 0.6 mm & \hphantom{-}67.2 $\pm$ 5.9 $\mu$s\\
\botrule
\end{tabular}
\end{minipage}
\end{center}
\end{table}

\subsection{Connecting MKEA Displacement to Mk-Pt \dUT\ Discontinuity}
\label{Anl:EstimatedShift}
With the observation from Section \ref{Anl:GNSS} that there was a coseismic displacement of the MKEA station, we now explore whether the magnitude and direction of that shift can account for the discontinuity in the Mk-Pt \dUT\ residuals. The expected discontinuity of \dUT\ residuals due to a shift in a participating station's position is calculated as the discrete total differential of \dUT\ with respect to the three coordinate axes of the ITRF, given as
\begin{equation}
\label{dUT1GNSScalc}
    \Delta(\text{\dUT})_{expected} = \sum_{A = X,Y,Z} S_A \Delta A
\end{equation}
where the $S_A = \partial (\text{\dUT})/\partial A$ are the sensitivities of \dUT\ to station position shifts and the $\Delta A$ are the station position shifts in each orthogonal coordinate.

Due to the proximity of the MKEA and MK-VLBA stations, we make the assumption that the station position shift of MK-VLBA due to the earthquake is the same as the displacement of the MKEA station. In this way, we already have the station position shifts of the MK-VLBA station.

Now we assess the sensitivity of the estimated \dUT\ value to offsets in a station's position relative to its presumed value. If a station was in a different place from the fixed position used in the calculation of the theoretical delay, some combination of free parameters in the least squares adjustment would have to deviate from their true values to accommodate that error. The clock parameters would be minimally affected as would the zenith wet delay parameters. To minimize the difference between the observed delay and computed theoretical delay, the model fit would ``rotate'' the earth, resulting in a shift in the value of \dUT.

To determine how much the estimate of \dUT\ would change due to a change in a station position that is unaccounted for, rather than trying to mimic the reality and develop a method of correctly altering the observed delays, we take advantage of the fact that real observed delays with simulated model station position shifts, and therefore artificial theoretical delays, provide the correct magnitudes of the sensitivities, just with the opposite sign, as we now demonstrate.

For a given observation of a source by the participating stations, the observed minus theoretical delay residual, retaining only the first term, is
\begin{equation}
    \Delta \tau = -\bf{Q}(\bf{b} - \bf{b_0}) \cdot \bf{s}
\end{equation}
where $\bf{Q}$ is the transformation matrix from terrestrial to celestial reference frame coordinates (the product of rotations for precession, nutation, UT1, and polar motion), $\bf{b}$ is the true baseline vector, $\bf{b_0}$ is the a priori baseline vector, and $\bf{s}$ is the unit vector in the source direction. Here, $\bf{Q}$ is defined to include the factor of the inverse of the speed of light ($1/c$). Shifting one station position results in a baseline change, $\Delta\bf{b}$. Adding a contribution to the actual delay corresponding to a change of $\Delta\bf{b}$ in the baseline vector
\begin{align*}
    \Delta \tau_1 &= -\bf{Q}([\bf{b} +\Delta \bf{b}] - \bf{b_0}) \cdot \bf{s} \\
        &= -\bf{Q}(\bf{b} - \bf{b_0} + \Delta \bf{b}) \cdot \bf{s} \\
        &= \Delta \tau - \bf{Q}(\Delta \bf{b}) \cdot \bf{s}.
\end{align*}
Conversely, modifying the theoretical delay with the same change $\Delta\bf{b}$
\begin{align*}
    \Delta \tau_2 &= -\bf{Q}(\bf{b} - [\bf{b_0} + \Delta\bf{b}]) \cdot \bf{s} \\
        &= -\bf{Q}(\bf{b} - \bf{b_0} - \Delta\bf{b}) \cdot \bf{s} \\
        &= \Delta \tau + \bf{Q}(\Delta \bf{b}) \cdot \bf{s}.
\end{align*}
Apparently, the modified delay residuals $\Delta \tau_1$ and $\Delta \tau_2$ are different from the original delay residual $\Delta \tau$ by the same magnitude, but with opposite signs. In the estimation process, $\Delta\tau$ is a linear combination of residuals due to each of the estimated parameters. The part of the total residual due to \dUT\ is directly proportional to the offset of \dUT\ from its a priori value. Thus, changes in \dUT\ due to identical station shifts applied to the true and assumed station positions will correspondingly be of the same magnitude but also with opposite signs.

After establishing control values of \dUT\ for hundreds of Kk-Wz and Mk-Pt Intensive sessions using the unaltered model station position, we estimate \dUT\ again for the same sessions after shifting, in turn, the model station position for a station in the Intensive network by +100 mm in the X, Y, and Z directions of the ECEF Cartesian coordinates of the ITRF. For each direction shift, station, and session, we then calculate the difference of the \dUT\ values estimated from the altered and unaltered station positions, divide it by the scale factor of 100 mm, and reverse the sign. The sensitivity is the mean of these values with the sample standard deviation reported as the uncertainty.

\begin{table}[ht]
\begin{center}
\begin{minipage}{76mm}
\caption{Sensitivities of \dUT\ to changes in the a priori station positions for MK-VLBA and PIETOWN in the Mk-Pt VLBA Intensives, and for KOKEE and WETTZELL in the Kk-Wz IVS Intensives. All units are $\mu$s/mm.}
\label{tab:sensitivity}
\begin{tabular}{@{}l|c|c@{}}
\toprule
Coord & Mk & Pt \\
\midrule
X & -0.5 $\pm$ 0.47
  & \hphantom{-}0.5 $\pm$ 0.47 \\
Y & -3.3 $\pm$ 0.31
  & \hphantom{-}3.3 $\pm$ 0.31 \\
Z & -4.4 $\pm$ 0.45
  & \hphantom{-}4.4 $\pm$ 0.45 \\
\midrule
  & Kk & Wz \\
\midrule
X & \hphantom{-}0.4 $\pm$ 0.06
  & -0.4 $\pm$ 0.06 \\
Y & -1.3 $\pm$ 0.09
  & \hphantom{-}1.3 $\pm$ 0.09 \\
Z & -0.1 $\pm$ 0.12
  & \hphantom{-}0.1 $\pm$ 0.12 \\
\botrule
\end{tabular}
\end{minipage}
\end{center}
\end{table}

For the Mk-Pt baseline, the sensitivities for a shift in the a priori station position of MK-VLBA and PIETOWN are shown in Table \ref{tab:sensitivity} along with the values for KOKEE and WETTZELL from the Kk-Wz baseline. As would be expected, the sensitivities of the PIETOWN station are the exact opposite of those of MK-VLBA, and likewise for KOKEE and WETTZELL.

Evaluating Equation \ref{dUT1GNSScalc} with the MKEA station displacements, $\Delta A$, from Table \ref{tab:MKEA_disp} (as proxies for the shifts of MK-VLBA) and the MK-VLBA sensitivities, $S_A$, from Table \ref{tab:sensitivity}, we find that the expected shift of the Mk-Pt \dUT\ residuals due to the shift in the MK-VLBA position is 
\begin{equation}
    \label{eqn:MKEA_dUT1_shift}
    \Delta(\text{\dUT})_{expected} = 67.2 \pm 5.9\ \mu s.
\end{equation}
The determination of the sensitivities and the station displacement as measured through GNSS are completely independent, so the errors are propagated assuming uncorrelated Gaussian uncertainties.

\section{Discussion}
\label{Discussion}

\subsection{Accounting for the \dUT\ Discontinuity}
\label{Disc:Discontinuity}

There are several potential causes of the discontinuity in the \dUT\ residuals measured by the Mk-Pt VLBA Intensives. With numerous inputs to the geometric model, a change in the reality of any of them (e.g., polar motion or celestial pole offsets) not reflected in the model assumptions could have an impact on the measurement of \dUT. However, the anomaly is only visible on data from the Mk-Pt baseline series, so the issue has to be associated with the participating stations, not the a priori EOPs. A one-time change in the observing stations' electronics could also have had an impact, but no such changes were detected in the other routine uses of the VLBA stations. Furthermore, a discontinuity in \dUT\ like this has been seen before in the IVS Intensive series that included the 32 m station at Tsukuba, Japan (TSUKUB32). This was traced to the effects of the 2011 Tohoku earthquake that resulted in a relatively large movement (at the level of a few meters) of the TSUKUB32 station \citep{macmillan_effects_2012}. All these factors strongly suggest that either the MK-VLBA or PIETOWN stations had suddenly moved. It is known that the PIETOWN station is not simply following the motion of the tectonic plate it sits on, but has a non-linear component to its long-term positional trajectory \citep{petrov_precise_2009} which potentially could have had a sudden shift. However, the coincidence of the K\={\i}lauea eruption and associated earthquake with the \dUT\ discontinuity pointed suggestively at MK-VLBA as the station that had moved and caused the jump in the \dUT\ measurements. 

To test this hypothesis we evaluated the size and direction of the jump seen in the VLBI data and developed a methodology of converting station position time series from co-located GNSS stations into an expected shift of the \dUT\ value. The two methods of calculating/estimating the \dUT\ displacement do comport with each other within a 3$\sigma$ threshold. The difference of the values reported in Section \ref{Anl:VLBI:MkPt_disp} (Equation \ref{eqn:MkPt_dUT1_shift}) and Section \ref{Anl:EstimatedShift} (Equation \ref{eqn:MKEA_dUT1_shift}) is 8.5 $\pm$ 7.5 $\mu$s, which is 1.14 $\sigma$.

There are many sources of noise that can contribute to the difference between the values produced by the two methods. As discussed in Section \ref{Anl:VLBI:MkPt_disp}, the selection of the outlier elimination threshold has an impact on the calculated discontinuity. The choice of the time to calculate the discontinuity also plays a small role in the final calculation. In this case, that time for both the VLBI and GNSS series was the same, but it was at the closest midnight epoch to the moment of the earthquake, not at the moment of the earthquake itself. The uncertainties of the measurements certainly capture some of these effects, but the systematic errors are likely underrepresented. Furthermore, the assumption that the position shift experienced at MKEA is identical to that experienced at MK-VLBA may be imperfect, and thus contribute to the discrepancy, even though the two stations are very close. Regardless, the results of Section \ref{Analysis} are consistent with the hypothesis that the station position displacement is what caused the value of \dUT\ to shift.

After such a station displacement event, VLBI global solutions are able to solve for the station position before and after the earthquake and thus also produce values for the position shift. However, in order to calculate this shift, several sessions involving the station in question must have been observed, which usually takes several months. For some stations, in addition to the initial station coseismic displacement, a post-seismic deformation model is required to account for non-linear station motion as the station settles into a new equilibrium \citep[see][Section 3.4]{altamimi_itrf2014_2016}.

In the case of the 2018 displacement of the MK-VLBA station, the \dUT\ residuals and the GNSS data are consistent with the event being a simple discontinuity, with no post-seismic deformation. This is supported by the results of the ITRF2020 analysis \footnote{\url{https://itrf.ign.fr/ftp/pub/itrf/itrf2020/ITRF2020-psd-vlbi.snx}}. Furthermore, a recent VLBI global solution from the USNO VLBI Analysis Center, \texttt{usn2022a}, calculates a position of the MK-VLBA station before and after May 4, 2018. The solution did not call for modeling a post-seismic deformation. The MK-VLBA station position shift from the solution is 6.8 $\pm$ 0.9 mm in X, -5.1 $\pm$ 1.0 mm in Y, and -13.0 $\pm$ 0.9 mm in Z, or 15.5 $\pm$ 0.9 mm slightly down and to the south east. This translates to an expected discontinuity in the Mk-Pt \dUT\ residuals of 70.6 $\pm$ 8.6 $\mu$s. Though these station position shifts are slightly different from those determined for the MKEA station in Section \ref{Anl:GNSS} (more so in X than in Y or Z), the total magnitude of the shifts of both MKEA and MK-VLBA and the corresponding expected induced discontinuity in the UT1$-$UTC value are similar. These measurements of the displacement of the MKEA and MK-VLBA stations are consistent in displacement direction with analysis of interferometric synthetic aperture radar (inSAR) data which concludes that the south-eastern flank of Maunakea (which is $\sim$90 km north west of the earthquake epicenter) slipped down and to the south east in the direction of K\={\i}lauea as a result of $\sim$5 m of fault slip to the south east at the site and time of the earthquake \citep{neal_2018_2019}.

The reality that an unforeseen station position shift can change the measurement of \dUT\ in low latency single-baseline Intensive measurements has some important consequences. Any station displacement would have an immediate effect on the measurement of \dUT. For Intensive series that are being used operationally to inform the global awareness of \dUT, any position shift needs to be detected and corrected quickly. The time between VLBI global solutions, which could be used to update the station a priori values, is at least a few months. The latency of detections of position shifts of GNSS stations is a few hours or days. This work, building on prior work by \citet{macmillan_effects_2012}, shows that the \dUT\ offset can largely be corrected by using the GNSS position shift in conjunction with the baseline's sensitivity of \dUT\ to a change in the position of the participating stations. The correction can be done using sensitivities for each affected station in a single baseline for all sessions, or the sensitivities could be determined on a session by session basis using the technique employed in Section \ref{Anl:EstimatedShift}. Regardless of how exactly the correction is determined, it can be calculated and applied quickly, supporting the low latency measurement of \dUT\ while the updated station position is determined more accurately from a global solution.

\subsection{Mk-Pt Intensive Series Characteristics}
\label{Disc:MkPtCharacteristics}

\begin{table*}[ht]
\begin{center}
\begin{minipage}{\textwidth}
\caption{Descriptive statistics of the Kk-Wz IVS Intensives and the three timeframes of the Mk-Pt VLBA Intensives with different session durations.}
\label{tab:stats}
\begin{tabular}{l|c|c|c|c|c}
\toprule
Intensive & Number of & Mean Number of            & Mean Formal     & Residual       & NWE Residual   \\
Series    & Sessions  & Obs per Session  & Error ($\mu s$) & WRMS ($\mu s$) & WRMS ($\mu s$) \\
\midrule
Mk-Pt (45 min) & 1012 & 28.3 & 22.0 & 36.2 & 31.8 \\
Mk-Pt (90 min) &  849 & 55.2 & 14.3 & 33.5 & 23.5 \\
Mk-Pt (60 min) &  142 & 41.7 & 20.5 & 31.7 & 28.6 \\
\midrule
Kk-Wz (60 min) & 1561 & 17.4 & 12.5 & 13.0 & 12.5 \\
\botrule
\end{tabular}
\end{minipage}
\end{center}
\end{table*}

By using different a priori station positions through time that account for station displacements, the residuals of Intensive series, computed with the method described in Section \ref{Anl:VLBI}, no longer contain any \dUT\ discontinuities, and thus provide another set of data with which to evaluate the Mk-Pt series. As mentioned in Section \ref{Obs:MkPt}, the session duration of the Mk-Pt Intensives changed over the course of the series. With the increased number of scans in longer sessions, it would be expected that the session formal error would go down, and perhaps so would the scatter of the residuals as quantified by the weighted root mean square (WRMS). Using the \dUT\ values of the Intensives calculated in the \texttt{usn2022a} solution, which accounts for the MK-VLBA station displacement, we separately examine the periods when the Mk-Pt VLBA Intensives had a duration of 45 minutes (February 15, 2013--February 23, 2017), 90 minutes (February 24, 2017--July 31, 2020), and 60 minutes (August 1, 2020--April 29, 2021), calculating the WRMS of the residuals and the mean formal error of the sessions for each time frame. We also calculate these values for the Kk-Wz baseline over the whole time span (February 15, 2013--April 29, 2021) for reference. The resulting statistics are shown in Table \ref{tab:stats}.

As expected, for the Mk-Pt baseline, the mean formal error decreases with increasing session duration. The WRMS of the residuals behaves similarly, though the 60 minute sessions have a lower residual WRMS than the 90 minute sessions. This is likely due to the dissimilarity of the number of sessions used in the calculation of the statistic. If the formal errors estimated for each session properly captured all of the systematic error, one would expect that the WRMS of the residuals would be equal to the mean formal error. As the values in the table show, the WRMS is consistently larger than the mean formal error, indicating an uncaptured systematic error contribution. Even for the Kk-Wz series, the WRMS of the residuals are slightly higher than the mean formal error.

In addition to calculating the \dUT\ discontinuity in the Mk-Pt VLBA Intensive series, the application of the NWE non-parametric regression highlights the presence of periodic structure in the residuals, visible in Figure \ref{fig:doublefit}. A similar structure might also be present in the Kk-Wz IVS Intensive series. This structure likely represents a part of the uncaptured systematic error of the reported formal errors that is causing the elevated WRMS in both series. The lower magnitude of the sensitivity of \dUT\ measurements from the Kk-Wz Intensives to changes in KOKEE and WETTZELL station positions relative to the sensitivity of the MK-VLBA and PIETOWN stations is a possible reason for the difference in amplitude of the periodic structure and requires additional investigation.

To provide a numerical understanding of the impact of the systematic errors in the \dUT\ residuals, we additionally calculate and report in Table \ref{tab:stats} the WRMS of the residuals of the \dUT\ time series after subtracting off the NWE regression function calculated at the epochs of all of the sessions in each time frame. For the Kk-Wz series, this results in a value of the WRMS of the ``corrected'' residuals that is identical to the mean formal error. For the Mk-Pt series segments, while the WRMS is lowered, the values remain higher than the mean formal errors, suggesting that the mean formal errors are under-estimated.

\section{Conclusion}
\label{Conclusion}
In this work we examined the effects of the $M_w$ 6.9 Hawai`i earthquake of May 4, 2018 on the MK-VLBA VLBI station and the MKEA GNSS station. From the analysis it is clear that MKEA had a coseismic displacement, which similarly affected the co-located MK-VLBA station. After calculating the sensitivity of the Mk-Pt Intensive measurement of \dUT\ to a shift in the MK-VLBA position, we calculated the expected discontinuity in \dUT\ from the observed MKEA coseismic displacement. The resulting value of 67.2 $\pm$ 5.9\ $\mu$s is consistent at the 1.14 $\sigma$ level with the \dUT\ discontinuity of 75.7 $\pm$ 4.6\ $\mu$s measured in the Mk-Pt \dUT\ series, very firmly connecting the coseismic station displacement to the discontinuity in the Mk-Pt \dUT\ residuals. Reinforcing that result is that the expected discontinuity in the \dUT\ residuals based on the \texttt{usn2022a} global solution station displacement is 70.6 $\pm$ 8.6 $\mu$s, which is 3.4 $\pm$ 10.4 $\mu$s (0.33 $\sigma$) from the expected discontinuity from the MKEA displacement and 5.1 $\pm$ 9.7 $\mu$s (0.52 $\sigma$) from the calculated discontinuity in the Mk-Pt residuals. Though the global solution was able to correct for the impact of the earthquake on Intensive measurements of \dUT, it was roughly a year before the USNO VLBI Analysis Center could make this correction.

Intensives monitor the vital EOP \dUT, providing a daily measurement with latency generally under 24 hours. Having a co-located GNSS station with a VLBI station participating in Intensives ensures that there is a mechanism to identify and correct any station displacements relatively quickly. With a simple discontinuity, an offset could be applied to all measurements of \dUT\ after the jump. However, in the case of an event that exhibits post-seismic deformation, the GNSS station position series would provide a means of correcting the Intensive series within a few days or weeks of the displacing event. Without a co-located GNSS station, corrections to the Intensive series from VLBI global solutions alone could take months, rendering that baseline unable to provide an accurate measurement of \dUT\ in low-latency Intensives. In order to maintain the needed frequent and low-latency measurement of the phase of the Earth's rotation from Intensive sessions, the international community would benefit greatly from ensuring that there is a GNSS station co-located with any VLBI station involved in these sessions.

In addition to allowing for a calculation of the discontinuity in the Mk-Pt \dUT\ residuals, the application of the NWE smoothing technique to the residuals reveals an oscillatory feature with a roughly semi-annual period. This feature is certainly present in the Mk-Pt series and is potentially present in the Kk-Wz series as well. Comparison of the mean formal error and the dispersion of the \dUT\ residuals indicates that there are systematic errors in the \dUT\ measurements that are not accounted for in the Mk-Pt Intensive series formal errors. The discovered oscillatory feature explains some of the excess dispersion relative to the mean formal errors.

Future work will focus on identifying and removing this ``wiggle'' from Intensive \dUT\ series. This work examined the sensitivity of \dUT\ to errors in modeled station position values, but only applied them to a single moment in time. With GNSS stations co-located with both the MK-VLBA and PIETOWN stations, a time series of corrections to the Mk-Pt \dUT\ series could be created. Furthermore, there can be errors in the a priori values for the polar motion parameters and the celestial pole offsets as well. Determining the sensitivities of any Intensive series to these errors may also shed light on the source of the oscillation in the \dUT\ residuals. With a semi-annual periodicity in the oscillation, investigation of potential issues with the modeled solid Earth tides or tidal loading may also be warranted. Through these, and complementary, investigations, we hope to better understand the nature of the ``wiggle'' that we have detected in measurements of \dUT\ from the Mk-Pt VLBA Intensives.

\backmatter

\bmhead{Acknowledgments}
Thanks are due to Julien Frouard of the USNO for his assistance with the application of the Nadaraya-Watson estimator, and also to Johnathan York of The University of Texas at Austin, Applied Research Laboratory and Sharyl Byram of the USNO for advice on matters related to the GNSS stations. The authors acknowledge use of the Very Long Baseline Array under the US Naval Observatory's time allocation. This work supports USNO's ongoing research into the celestial reference frame and geodesy. The VLBA is an instrument of the National Radio Astronomy Observatory which is a facility of the National Science Foundation operated under cooperative agreement by Associated Universities, Inc.

\section*{Declarations}
\bmhead{Authors' contributions}
CD and MJ designed the study. DM developed the sensitivity calculation method which was implemented by CD. CD performed the analyses, executed the calculations, prepared the figures, and drafted the manuscript. All authors interpreted the results and provided input that shaped the analysis and manuscript.

\bmhead{Availability of data and materials}
All data products utilized in this analysis are in the public domain. Links to their sources are provided in the text.

\bmhead{Competing interests}
The authors declare that they have no conflicts of interest.

\bibliography{main}

\end{document}